\documentclass[fleqn,10pt]{wlscirep}
\usepackage{url}

\title{ Ultrafast Relativistic Electron Nanoprobes}

\author[1]{F. Ji}
\author[2,3]{D. B. Durham}
\author[2,3]{A.M. Minor}
\author[4]{P. Musumeci}
\author[4]{J. G. Navarro}
\author[1,*]{D. Filippetto}

\affil[1]{Accelerator Technology and Applied Physics Division, Lawrence Berkeley National Laboratory, One Cyclotron Road, Berkeley, California, 94720, USA}
\affil[2]{National Center for Electron Microscopy, Molecular Foundry, Lawrence Berkeley National Laboratory, One Cyclotron Road, Berkeley, California, 94720, USA}
\affil[3]{Department of Materials Science and Engineering, University of California, Berkeley, California, 94720, USA}
\affil[4]{Department of Physics and Astronomy, University of California, Los Angeles, California, 90095, USA}

\affil[*]{dfilippetto@lbl.gov}


\begin{abstract}

One of the frontiers in electron scattering is to couple ultrafast temporal resolution with highly localized probes to investigate the role of microstructure on material properties. Here, taking advantage of the unprecedented average brightness of the APEX electron gun providing relativistic electron pulses at high repetition rates, we demonstrate for the first time the generation of ultrafast relativistic electron beams with picometer-scale emittance and their ability to probe nanoscale features on materials with complex microstructures. At the sample plane, the APEX beam is tightly focused by a custom in-vacuum lens system based on permanent magnet quadrupoles, and its evolution around the waist is tracked by a knife-edge technique, allowing accurate reconstruction of the beam shape and local density. We then use the focused beam to characterize a Ti-6 wt\% Al polycrystalline sample by correlating the diffraction and imaging modality, showcasing the capability to locate grain boundaries and map adjacent crystallographic domains with sub-micron precision. This work provides a new paradigm for ultrafast electron instrumentation, demonstrating the ability to generate relativistic beams with ultrasmall transverse phase space volumes enabling novel characterization techniques such as relativistic ultrafast electron nano-diffraction and ultrafast scanning transmission electron microscopy.
\end{abstract}
\begin{document}

\flushbottom
\maketitle

\thispagestyle{empty}


Since the discovery of the particle-wave duality \cite{Nobel}, electrons have been extensively used to probe matter at atomic scales. Owing to their very short (sub-{\AA) wavelength and large scattering cross section compared to X-rays, electron diffraction and imaging are today well established techniques for structure determination. More recently, the advent of ultrafast lasers sparked the development of intense ultrashort electron sources which, in turn, paved the way to a new generation of time-resolved electron scattering techniques such as ultrafast electron diffraction and microscopy (UED/M)\cite{Zewail,SciainiMiller,king_JAP2005}. These are now capable of probing atomic-scale structural dynamics with femtosecond-scale temporal accuracy.

Recent developments in this field include the introduction of methods and technology common in particle accelerator science. Radio frequency (RF) based electron sources have been successfully used for generating few-femtosecond electron probe beams \cite{maxson2017,zhao2018} and for gathering information about ultrafast structural changes in solids and gases\cite{SLAC_first,yang2016}. Here, electrons are generated and rapidly accelerated to relativistic energies by using high accelerating gradients, increasing the maximum achievable electron current density \cite{BazarovPRL,filippetto2014} and minimizing the temporal broadening caused by Coulomb repulsions and initial energy bandwidth, which are the main challenges for low-energy electron sources. 

Notwithstanding this significant progress, ultrafast electron-based instrumentation is still far from reaching spatial resolution similar to what can be achieved in static electron microscopes. At low energies, setups using tip-based photoemission guns in standard transmission electron microscope columns can take advantage of very small source areas and DC accelerating fields \cite{ropers2014,Arbouet2017}. On the other hand, all pump-probe studies with MeV electrons up to date have so far been limited to systems with long-range order in the tens of micrometers or more \cite{SLAC_UED,BNL_UED}, as a consequence of the limited average transverse beam brightness $B_{n,av}$ available from the source. This quantity, defined as the number density of electrons in transverse phase space (i.e. per unit solid angle and unit area, also called 4D emittance), sets a limit to how much a beam can be focused before its intrinsic divergence overwhelms any deflections due to scattering from the sample. For a pulsed electron source, the most direct way to improve the average transverse brightness is by increasing the repetition rate, which, for very high gradient RF guns, is typically limited to few hundred Hz . 

In the present study we use relativistic ultrafast electron pulses to map structural variations in microstructured materials, demonstrating for the first time a relativistic UED probe with nanoscale spatial resolution. The setup benefits from a unique electron beam source \cite{sannibale2012} capable of generating a $B_{n,av}$ more than 2 orders of magnitude beyond previous setups.
First, we demonstrate lateral focusing of the electron beam down to the nanometer scale. The electron beam is opportunely collimated and then injected into the experimental chamber for final focusing using custom in-vacuum permanent-magnet lenses. The unprecedented beam quality coupled with strong focusing required the development of a novel beam characterization technique, based on knife-edge scanning measurements complemented by a detailed data analysis. We obtained a full reconstruction of the beam transverse phase space evolution near the waist yielding transverse spot sizes smaller than 100 nm and normalized emittance in the sub-nanometer range.

We then show the potential for such beams in ultrafast nano-diffraction (relativistic nano-UED) and scanning transmission electron microscopy (relativistic USTEM) by recording high quality diffraction patterns and mapping grain orientation with sub-micrometer resolution, pinpointing a grain boundary with sub-spot size resolution. In this contest, we first characterize the ultrafast point-projection microscopy mode of the instrument (relativistic UPPM), obtaining a spatial resolution consistent with the knife-edge waist size measurements \cite{mullerPPM2014,bainbridgeUPPM2016,quinonez_UPPM2013}, and then elucidate the critical role of this modality in the context of the nano-UED experiments. UPPM in fact, is shown to provide key information which can be used for correlating the ultrafast structural dynamics data from nano-UED with static information retrievable with other techniques such as conventional TEM. 

\section*{Results}
\subsection*{Lateral squeezing of ultrafast electron beams}

\begin{figure}[]
\centering
\includegraphics[width=0.75\linewidth]{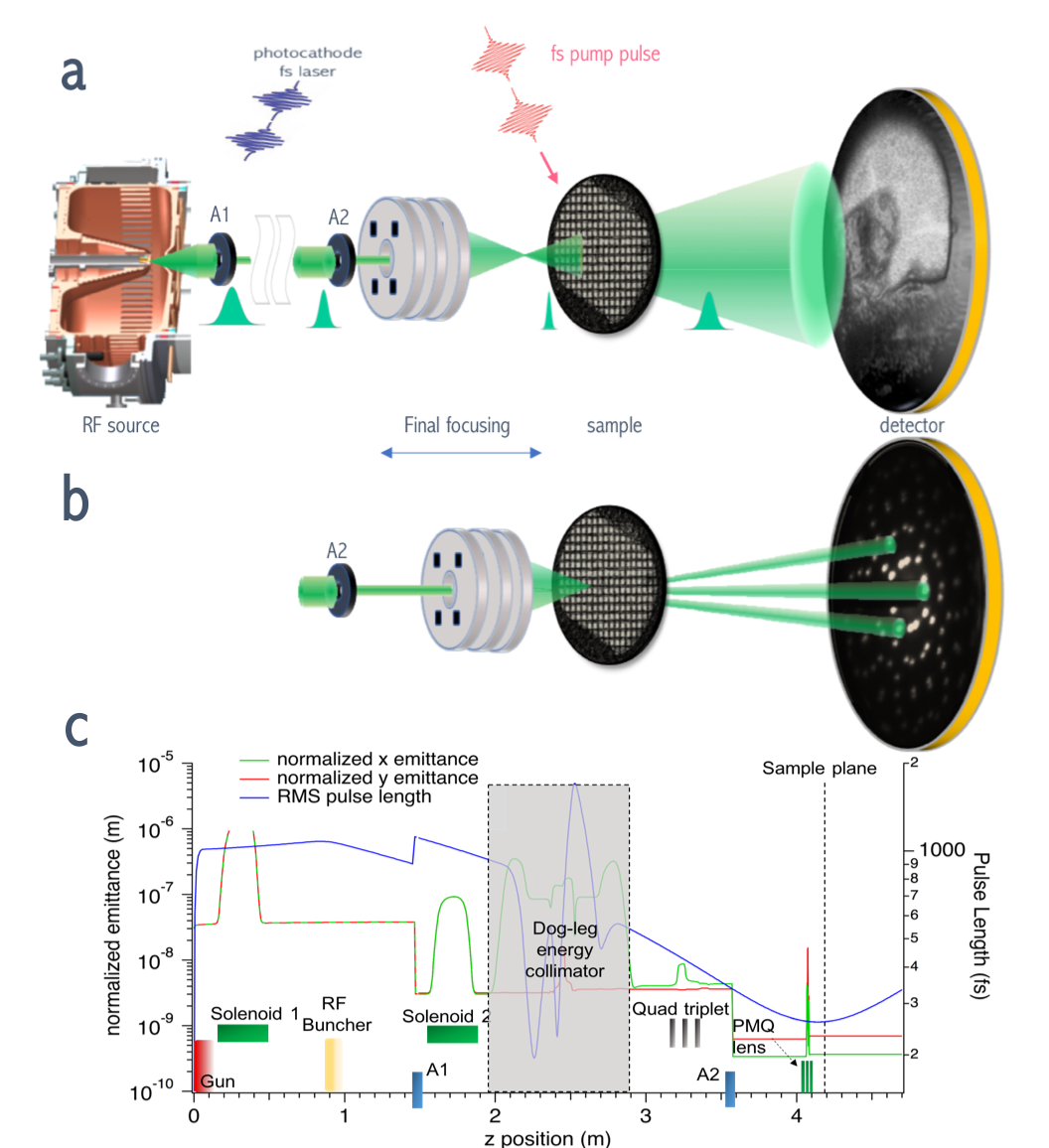}
\caption{Experimental setup. \textbf{a} Cartoon of the electron beamline for ultrafast nano-UED experiments. From left to right: a section of the radio-frequency electron gun showing the internal nose-cone shape maximizing the accelerating field along the electron beam trajectory; two apertures (A1 and A2) are then used to select electrons with low transverse momentum, the final focusing lens composed by permanent magnet quadrupoles focused the electrons which are then intercepted by a scintillator screen. Here the position of the lens is such that the waist is produced upstream the sample plane, producing a shadowgraph of the specimen. The green Gaussian waveform represents a qualitative behaviour of the beam temporal evolution. \textbf{b} The same schematic with the setup operating in diffraction mode, with coincident electron beam focus and sample planes \textbf{c} Electron beam dynamics simulations showing the behaviour of the electron beam emittance and pulse lenght throughout the beamline. The two apertures A1 and A2 decrease the electron beam emittance by about 1 order of magnitude each. At the same time, a negative energy-time correlation is imprinted on the electron beam by the radio-frequency buncher, which causes the beam to compress in the subsequent vacuum drift, reaching a minimum value at the sample plane.}
\label{fig:Exp}
\end{figure}

Nanoscience applications of ultrafast relativistic electron pulses pose stringent requirements on the electron beam properties. In the case of diffraction, resolution can be described using the resolving power $\mathcal{R}=R_{hkl}/\Delta R_{hkl}$ \cite{DCdesign_1953}, where $R_{hkl}$ is the distance of a specific diffraction point/ring from the zero-order beam while $\Delta R_{hkl}$ represents the minimum distance at which a second point/ring can be discriminated. The resolving power in diffraction is ultimately limited by the angular spread of the beam, $\sigma_{u'}$, which is inversely linked to the spot radius $\sigma_u$ at the waist by the beam normalized emittance (ie. $\epsilon_{n,u} = \sigma_u\sigma_{u'}$). To achieve a given resolving power at a particular spot size, the emittance requirement is $\epsilon_{n,u} = \frac{\lambda_c\sigma_u}{2d_{hkl}\mathcal{R}}$, where $\lambda_c$ is the electron Compton wavelength and $d_{hkl}$ is the inter-atomic separation distance. For example, to achieve $\mathcal{R}=10$ with a nanoscale beam ($\sigma_u=500~nm$) in a sample with an inter-atomic distance $d_{hkl}=2$~\AA, the required normalized emittance is $\epsilon_{n,u}=300~pm$. Such a value is more than one order of magnitude beyond the smallest emittance experimentally measured at present date in relativistic ultrafast electron beamlines \cite{UCLAnanoemittance}.

This work utilizes the High Repetition-rate Electron Scattering (HiRES) beamline, a recently developed UED/M instrument that employs a unique RF-based electron source to achieve ultrashort, low-emittance electron beams. The details and performance of HiRES are described elsewhere \cite{HiRESsims}.

In Fig.\ref{fig:Exp}a we show a cartoon of the apparatus, summarizing the beamline elements and functions relevant for this work. Short bursts of electrons are generated via photoemission at a 1 MHz rate, and instantaneously accelerated to relativistic energy of 735~keV via rapidly oscillating electromagnetic fields with an amplitude of $20$ MV/m~\cite{APEXgun}. They then travel through the 4.5~m long electron transport line, which performs spatial filtering, energy collimation and longitudinal compression before reaching the sample. In particular, an RF-based compression cavity (RF buncher) imparts a negative energy-time correlation to the electron distribution which results in beam temporal compression through vacuum dispersion. Figure~\ref{fig:Exp}a includes a schematic view of the electron pulse length evolution along the beamline. The optimal electric field amplitude depends on initial beam energy modulation and pulse length \cite{HiRESsims}, and it is carefully tuned so the pulse is shortest at the sample position (260~fs RMS in Figure~\ref{fig:Exp}c). 

The transverse electron beam properties are shaped at several points along the beamline. 
First, a stream of electron beam pulses with an average current of 60~nA is generated using a laser pulse with 300~fs full width at half maximum (FWHM) transversely focused to a 50 $\mu$m root-mean-square (rms) spot on the photocathode. The combination of the first solenoid and a collimating aperture with a fixed diameter of $500~\mu m$ (A1 in Fig.~\ref{fig:Exp}a) downstream from the source selects the particles with low transverse momentum, thereby filtering the transverse phase space. The transverse normalized emittance of the resulting 320~pA beam is about $3~nm$, measured by reconstructing the transverse phase space at the aperture position via TEM grid shadowgraph analysis \cite{MarxNavarro_prab}. 
Upstream of the experimental chamber, and after the dog-leg transfer line, the electron beam is spatially filtered again by a second aperture with variable diameter from 1~mm down to $10~\mu m$ (A2 in Fig.~\ref{fig:Exp}a) to reach sub-nanometer emittance values. The electron optics downstream of A1 (not shown) are tuned to modulate the transverse aspect ratio of the beam at A2 and, consequently, partition the four-dimensional emittance to create round or flat beam waists. Typical current values after the second collimator are in the range of 100-200~fA. 

Focusing relativistic electron beams to nanoscale spots requires strong confining magnetic fields. In this experimental work we explore the use of an in-vacuum lens assembly composed by 3 permanent magnet-based focusing elements\cite{Nanoquadrupoles1,Nanoquadrupoles2} as a compact alternative to large solenoid lenses. 

We designed and fabricated quadrupole lenses (PMQ) with focusing gradients in excess of 100~T/m using Neodymium-based permanent magnets ($Nd_2Fe_{14}B$, remanence $B_r=1.25~T$). We then arranged them in a triplet configuration to achieve an overall focal length of $f_e=2.5~cm$ in both planes. Figure~\ref{fig:PMQs} shows a single quadrupole element of the focusing assembly, with four 3~mm-thick permanent magnets held together by a round aluminum holder with a bore aperture diameter of 4~mm. The relative longitudinal distance between midpoints of the 3 elements was optimized using particle tracking simulations \cite{GPT}. Figure~\ref{fig:PMQs} reports an example of simulation where a round input beam with an $50~\mu m$ rms size and 600~pm normalized emittance is focused down to 400 nm rms about 2.5~cm downstream the exit of the third focusing element, for relative distances between the quadrupoles of 5 mm and 6.5~mm respectively.

Precise transverse alignment and field measurements of the PMQ lenses are required to achieve the target focal length and beam size. The three quadrupole magnets were aligned using the pulse-wire technique to a common axis with a tolerance of $<15~\mu m$ ($<3~mm$ transverse kick at the detector) and their gradient was measured with an error better than $<1\%$ of the specification ($<5~mm$ error in longitudinal position of focal plane), within the acceptance tolerances set by a parametric simulation study of focal length variation and transverse dipolar kicks.

Figure~\ref{fig:PMQs}b shows the final in-vacuum mechanical configuration for the experiment. The flexure structure embracing each quadrupolar element was designed to allow such a precise transverse alignment of the magnet while minimizing longitudinal footprint. To allow compensation for small errors in focal length, an in-vacuum linear piezo-actuator was added to the system to adjust the longitudinal position of the third element during the experiment (labeled as PMQ-3). In addition, the entire triplet assembly can be translated longitudinally over 2~cm, and horizontally in and out of the beam path, while the sample holder can be moved horizontally and vertically. 

\begin{figure}[ht]
\centering
\includegraphics[width=0.75\linewidth]{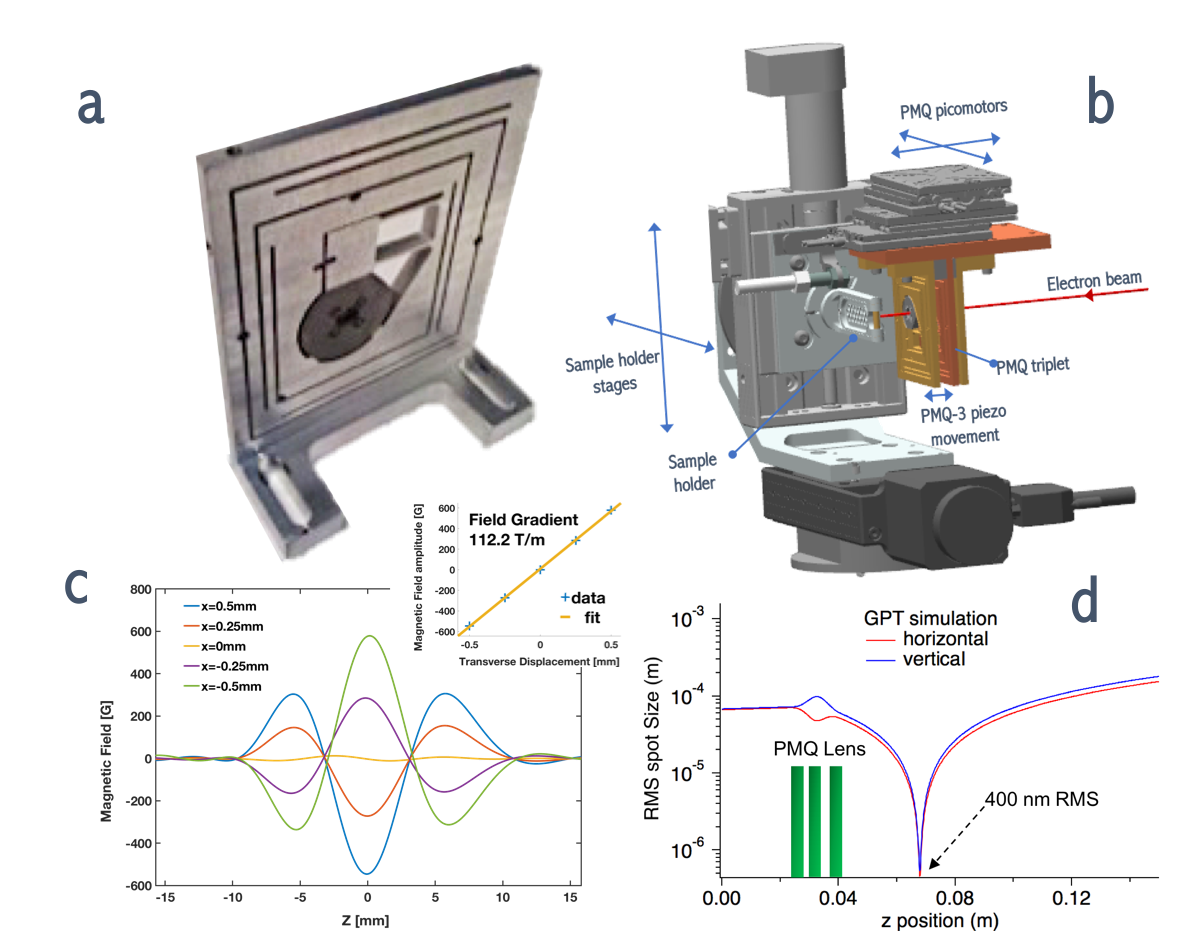}
\caption{Design and characterization of the final focusing lens system. \textbf{a} Single quadrupole element. An aluminum disk of 0.625 inch diameter holds 4 permanent magnets placed in quadrupolar configuration. The outer square flexure is specifically designed to minimize the footprint, and provide at the same time high precision control on transverse alignment of the magnetic element. \textbf{b} Three-dimensional view of the sample area. The sample holder can be moved horizontally and vertically with 100~nm precision. The focusing system can be moved longitudinally and horizontally with better than 10~nm precision. Additionally, the longitudinal position of the last quadrupole element can be adjusted with $sub-\mu m$ precision. \textbf{c} Transverse and longitudinal alignment of magnetic elements. The plot shows the magnetic field magnitude moving along the focusing system. With no offset (x=0) the peak measured field was below 15~G, corresponding to a transverse misalignment error of less than $14~\mu m$. The inset shows the measured focusing strength of one of the elements. \textbf{d} Beam dynamics simulation of a collimated electron beam entering the focusing system. Such simulations were performed using the actual measured gradient profiles, with the distances between the elements optimized to minimize the overall focal length of the system.}
\label{fig:PMQs}
\end{figure}

\subsection*{Experimental characterization of electron beam transverse phase space evolution near the waist}
\label{ssec:nanobeams}

\begin{figure}[ht]
\centering
\includegraphics[width=0.6\linewidth]{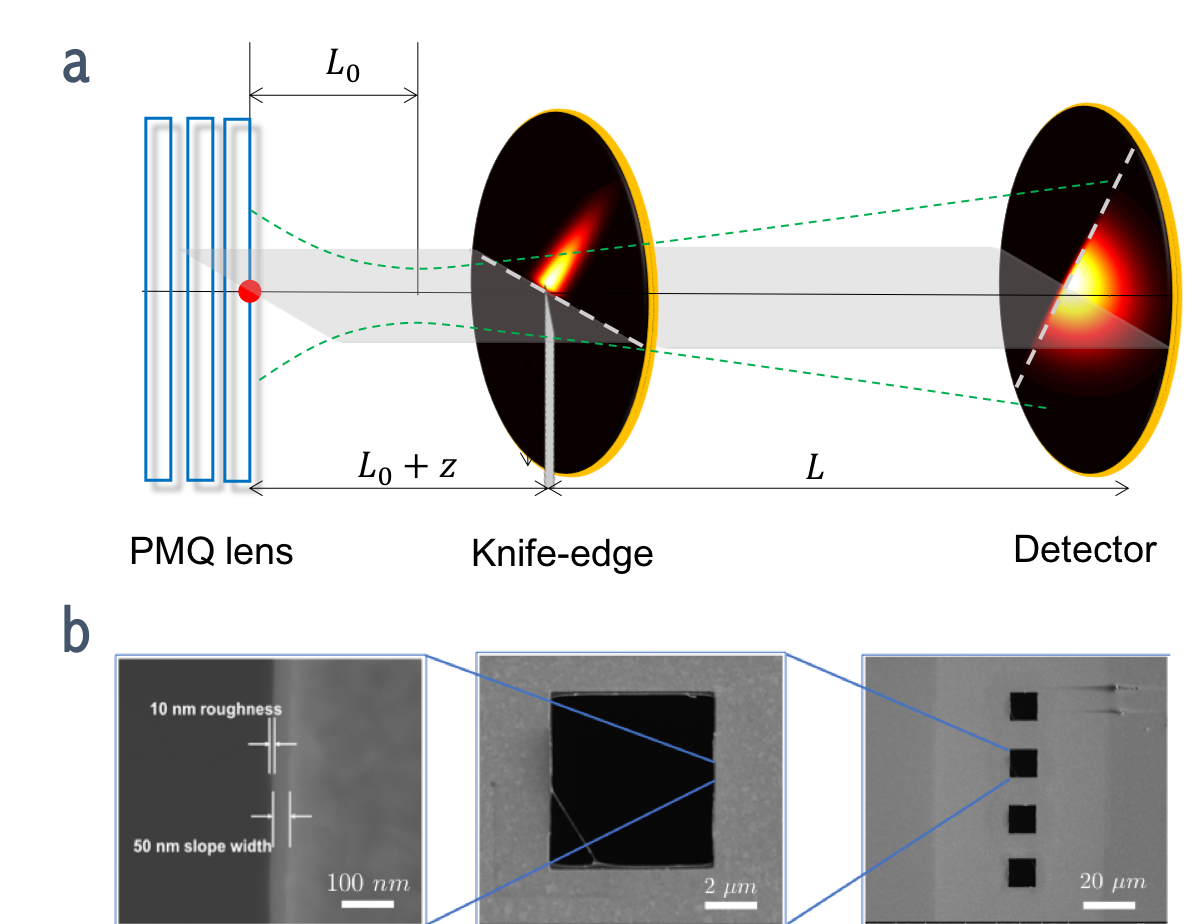}
\caption{Schematic of the measurement technique.\textbf{a} The electron beam is focused by the final lens and is then intercepted by a knife-edge target. The resulting beam image is collected at the detector. The position between the final lens and the target can be varied with $<$ 10~nm precision. \textbf{b} SEM of the knife-edge target. }
\label{fig:Technique}
\end{figure}

Measurement and control of relativistic electron beams with nanometer resolution is an active field of research. 
Recently, measurement of sub-micron electron beams has been reported using the beam-loss monitor signal generated by interaction of the electron beam with a nano-fabricated wire \cite{PSI_nanobeam}. In our setup we use a similar approach - a knife-edge target is inserted gradually into the beam along the horizontal and vertical direction - but we record the full beam image at the detector for each step (Fig.~\ref{fig:Technique}a-b). This allowed for a detailed analysis yielding the full phase space reconstruction, uncovering important correlations between the horizontal and vertical plane which would not be seen otherwise. The PMQ lens was moved along the direction of electron propagation, acquiring data at different longitudinal locations (Fig.~\ref{roundbeam}a and Fig.~\ref{roundbeam}b).

The knife-edge target used in the measurements is shown in Fig.~\ref{fig:Technique}b Focused ion beam (FIB) was used to mill $10 x 10~\mu m$ square holes from 30~nm SiN windows with 75~nm of gold deposited via thermal evaporation. SEM images of the square edges reveal a roughness of about 10~nm, together with a 50~nm-wide area with rapidly varying gold thickness

We perform a 10-parameter global fit on this dataset to reconstruct the coupled four-dimensional particle distribution in the canonical phase space $(x,p_x,y,p_y)$ \cite{KnifeEdge_emitt}, the related 4x4 second order beam matrix and its RMS volume $\epsilon_{n4D}=0.0144 (nm\cdot rad)^2$.  The evolution of the eigenvalues of the beam matrix in the \textit{xy} plane allow the determination of the position and size of the beam waists (Fig.~\ref{roundbeam}c), together with its rotation angle. The beam size minima were found to be $363$ and $609~nm$ at the specific longitudinal planes shown by the transverse sections in Fig.~\ref{roundbeam}d.

Asymmetric emittance and spot sizes can be achieved by changing the beam aspect ratio at the second aperture (A2 in Fig.\ref{fig:Exp}c) to take advantage of the dependence of the apertured beam transverse emittance on the angular divergence distribution before the aperture plane. By re-tuning the upstream quadrupoles to control this divergence, we were then able to generate beams with spot size aspect ratio up to 10 and minimum dimension at the focal point of $91~nm$ (Fig.~\ref{flatbeam}a-b). Such electron probes could be particularly useful in situations where high resolution is only needed in one dimension. 

An electron beam rotation in the transverse plane is observed and reported in Fig.~\ref{flatbeam}c. This is a result of an initial non-vanishing correlation between transverse planes, which can be due to various factors including an asymmetric initial electron beam distribution at the cathode coupled with rotation by upstream solenoid lenses and different focal lengths of the final PMQ lens in the two transverse directions. Once measured, and characterized, differences in PMQ focal lengths can be eliminated by adjusting the longitudinal position of the last PMQ quadrupole and by adding a skew-quadrupole compensation optics. It is worth noting that, after the subtraction of 90 degrees originated by the crossing of the horizontal axis by the beam and the consequent exchange of the minor and major axis of the beam ellipse (Fig.~\ref{flatbeam}a), the beam angle in space at the two waists position is off by about 3 degrees. Such small residual angle could be given by a small angular momentum introduced for example by a rolling alignment error of the PMQ quadrupoles. 

\begin{figure}[ht]
\centering
\includegraphics[width=0.98\linewidth]{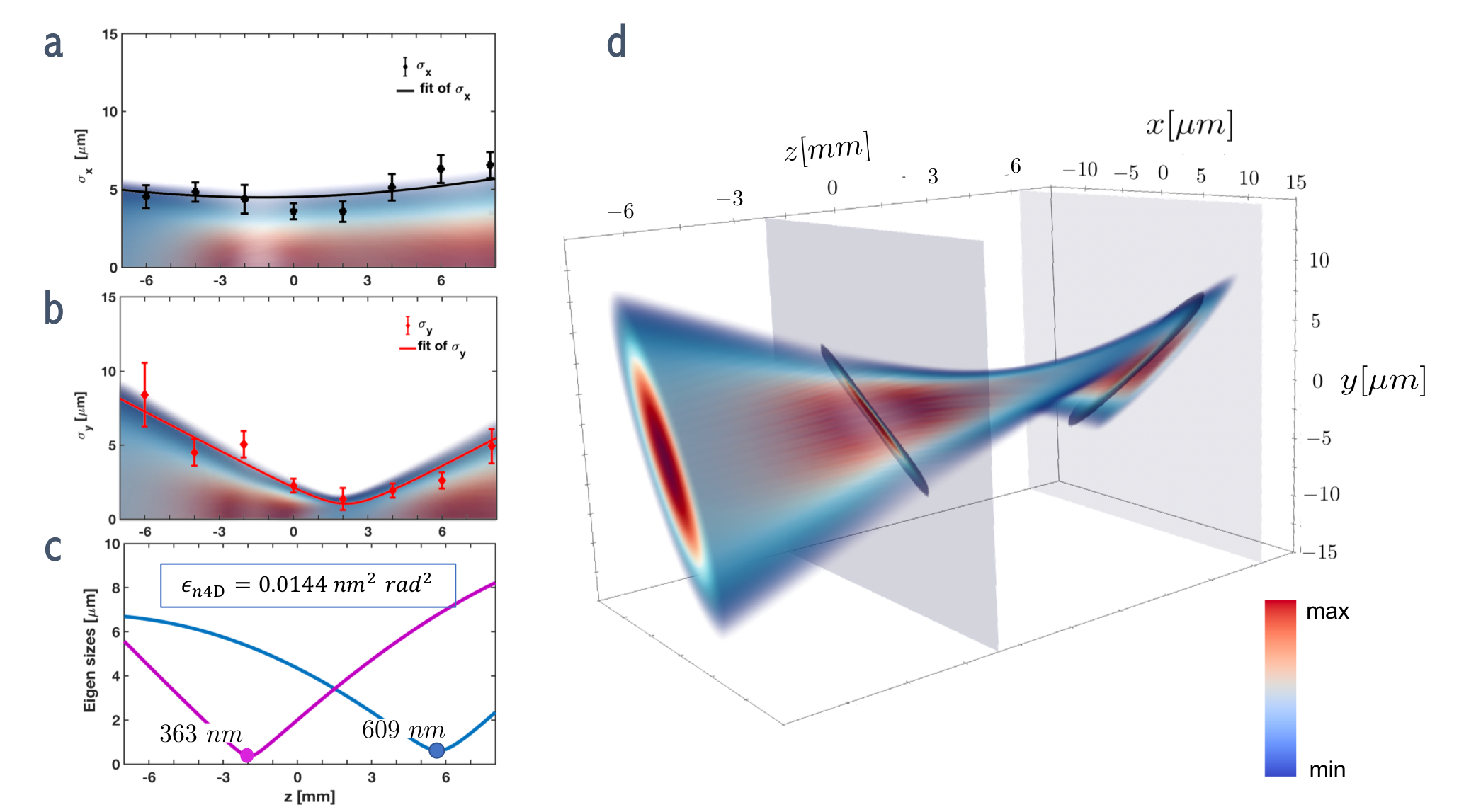}
\caption{Electron beam evolution around the waist. \textbf{a} Electron beam evolution projected on the horizontal axis. Black dots represent the experimental data, while the solid line shows the result of the global fit. Also reported in the plot is the density distribution of the beam, i.e. the projection of the volumetric rendering in \textbf{d} onto the horizontal axis. \textbf{b} Electron beam evolution projected along the vertical axis, similar to \textbf{a}. \textbf{c} Evolution of the spatial eigen sizes of the four-dimensional beam matrix. As the beam is rotating in space, a new diagonal matrix is found for each longitudinal position, revealing the beam orientation angle and beam size along diagonal directions. \textbf{d} Volumetric reconstruction of the electron beam density evolution. The two longitudinal slices shown represent the positions of electron beam waist.}
\label{roundbeam}
\end{figure}

\begin{figure}[ht]
\centering
\includegraphics[width=0.95\linewidth]{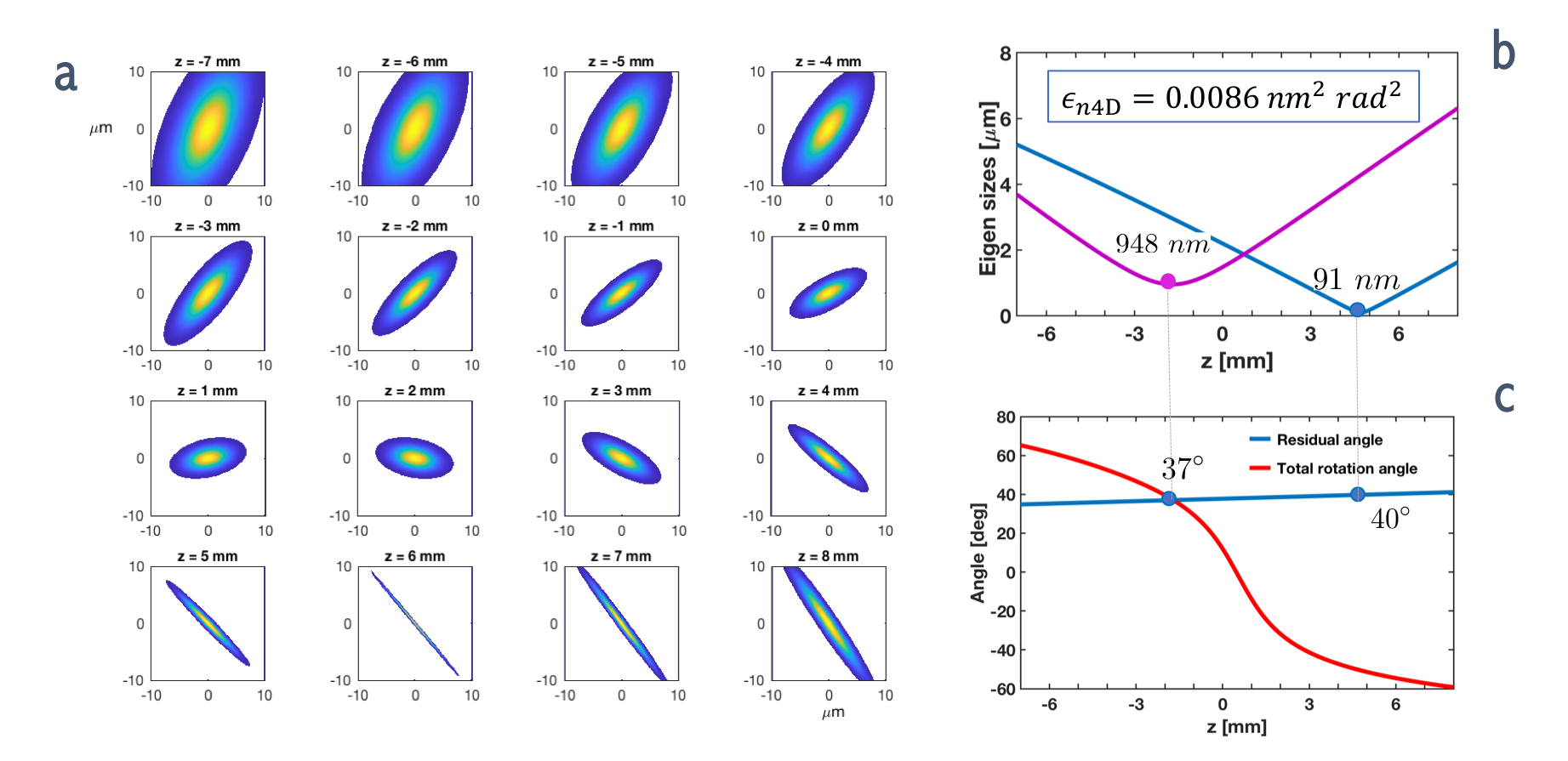}
\caption{Electron beam evolution around the waist for a beam with large aspect ratio. \textbf{a} Spatial evolution of the beam revealing the longitudinal position of the transverse waist and the rotation angle in the \textit{xy} plane. \textbf{b} Evolution of eigenvalues in the \textit{xy} configuration space, showing a  10:1 aspect ratio for the waist size, and a minimum of 91~nm. \textbf{c} Rotation angle of the associated eigenvectors as function of longitudinal position. A small residual angle is present after removing the $90^\circ$ geometric contribution due to the difference in the beam waist positions.}
\label{flatbeam}
\end{figure}

\subsection*{Femtosecond relativistic point-projection microscopy at the nanoscale} \label{ssec:PPIsection}

When the beam focus position is set upstream the sample plane, the instrument operates in imaging mode, performing ultrafast point-projection microscopy (UPPM). The focal plane is positioned upstream of the sample so the resulting image at the detector represents a magnified mass-contrast shadowgraph of the specimen. To study the resolution of our system in imaging mode, we extended the concept of Ronchi ruling to electron optics \cite{ronchi1923frange,osterberg1962evaluation}. We fabricated horizontal and vertical three-bar rulings with width and spacing ranging from 1.1 $\mu$m to 300 nm (Fig.~\ref{fig:PPI}a). Such rulings provide targets of known spatial frequency composition which can be used to determine the image contrast as a function of frequency, known as the contrast transfer function (CTF). The material is 50 nm of AuPd alloy sputtered onto a 30 nm SiN membrane, and the gaps were milled through with a focused Ga ion beam. With the electron beam focused in the configuration depicted in Fig.~\ref{roundbeam}, we formed an image of the target at the detector by accumulating ultrafast point-projection images at 1 MHz repetition rate for 1 second to obtain Fig.~\ref{fig:PPI}b. The target longitudinal position was chosen to minimize shear and stretching distortions in the final image (z=0 in Fig.~\ref{roundbeam}c).

From this image we determine the contrast transfer function of the instrument, shown in Figure~\ref{fig:PPI}c. We extracted contrast values from the three largest rulings at their fundamental frequency $f_0$ and $\frac{f_0}{2}$ (see Methods for detailed procedure), and we fit a Gaussian CTF. We verified this CTF by applying it to model gratings with the SEM-measured ruling dimensions to generate the profiles shown in the inset of Fig.~\ref{fig:PPI}c. These profiles reproduce the shapes of the overlaid measured profiles. From this fit, we find the spatial frequency resolution at 5\% contrast to be 0.725 $\pm$ 0.012 $\mu$m$^{-1}$ in X and 0.960 $\pm$ 0.017 $\mu$m$^{-1}$ in Y.

The UPPM resolution is set by the angular divergence of the electron probe, which in turn is defined by the beam size at its waist. Therefore we expect the point spread function (PSF) of the system to be closely related to the beam waist size (see Methods section). For the round beam case of Fig.~\ref{roundbeam}, we computed the PSF from our fit CTF and found its standard deviation to be $406\pm 7~nm$ and $539 \pm 9~nm$ respectively in the horizontal and vertical planes, in fair agreement with the minimum beam sizes measurements.

\begin{figure}[ht]
\centering
\includegraphics[width=0.8\linewidth]{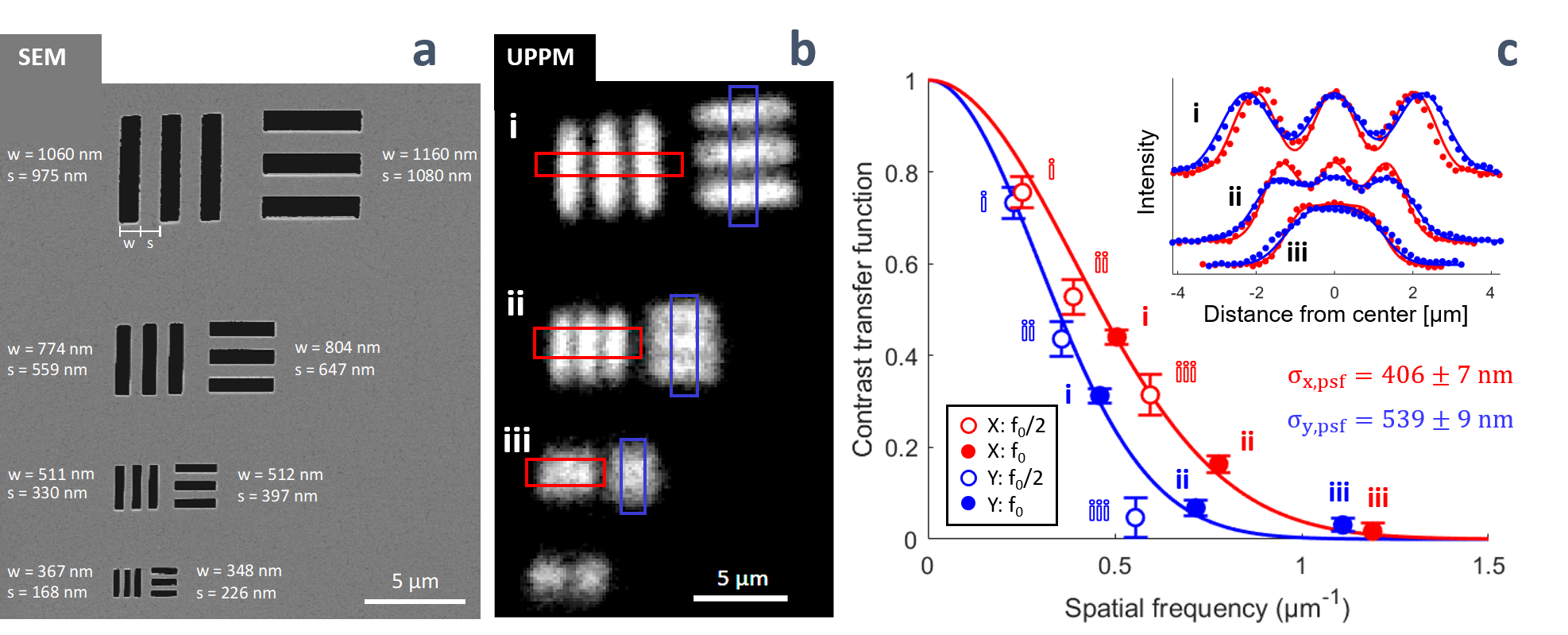}
\caption{Relativistic ultrafast point-projection microscopy calibration. \textbf{a} SEM image of the resolution target used to determine the contrast transfer function of the instrument. The Ronchi rulings are labeled with their dimensions as measured by SEM. \textbf{b} Ultrafast electron point-projection image of the target. Line profiles were extracted from the red and blue regions for resolution analysis. \textbf{c} Contrast transfer function of the instrument. Contrast values obtained from the measured line profiles at $f_0$, the fundamental frequency, and $\frac{f_0}{2}$ are plotted as points and fit with Gaussian CTFs (solid curves). The inset shows the measured ruling profiles (points) superimposed with model profiles (curves) computed by applying the best-fit CTF to step functions with the SEM-measured dimensions. For a detailed description of contrast transfer function determination, see the Methods section.
}
\label{fig:PPI}
\end{figure}

\subsection*{Scanning electron nano-diffraction with ultrafast relativistic electrons} \label{ssec:USTEMsection}

When focused at the sample plane, the low-emittance electron beam produces high quality diffraction patterns providing structure and orientation information at the nanoscale. We used the beam described in Fig.~\ref{flatbeam} to map grain orientation and boundaries in a hexagonal close-packed Ti-6 wt.\% Al ($Ti-6Al$ in the following) polycrystal. The sample was thinned by jet polishing to create a hole surrounded by ultrathin regions.  Electron backscatter diffraction (EBSD) in an SEM was used to create a reference map of the grain orientations near the hole (Fig.~\ref{fig:USTEM}a). Using the point-projection microscopy mode described above, we located and imaged the largest protrusion into the hole to determine the sample in-plane orientation (Fig.~\ref{fig:USTEM}b). We then focused the beam onto the protrusion and obtained a diffraction pattern indicating [0001] in that grain is nearly normal to the sample, matching the orientation determined using EBSD. 
%
%

We then demonstrated the ability of the ultrafast beam to locate a grain boundary with sub-diameter precision, by scanning the sample stage along one axis and acquiring a diffraction pattern at each 250 nm step. The orientation fraction, the fraction of electrons that pass through a grain, is $f_{j} = \frac{(I_{j}/I_{j}^o)}{\sum_{k} (I_{k}/I_{k}^o)}$; $I_{j}$ is the total intensity of a chosen set of diffraction spots due to grain j and $I_{j}^o$ is $I_{j}$ when the probe is entirely within grain j.
We compute $I_{j}$ by fitting symmetric 2D Gaussian functions to the selected peaks (circled in the figure) and summing their intensities. 
The orientation fractions of the two grains over the scan are shown in Fig.~\ref{fig:USTEM}c. 
The scanning interface over the beam produces a cumulative distribution function of the beam intensity along the scan direction. Each orientation fraction is thus expected to follow a Gaussian error function for a Gaussian beam. 
By fitting Gaussian error functions in the two directions, we identify the crossing point (the center of the interface) with sub-diameter precision: fitting error for the interface position is $\pm$ 15 nm standard deviation. This demonstrates the potential for precise nano-UED studies of interfaces.

Finally, we formed an ultrafast scanning transmission electron microscopy (USTEM) image by mapping the crystal orientation throughout a continuous bend in the sample over a 20 $\mu$m x 14 $\mu$m area in 2 $\mu$m steps (Fig.~\ref{fig:USTEM}d). We identified four on-zone orientations along this bend in the diffraction patterns, made possible by the nanoscale beam size. We approximate the orientation at each location to be an average of these four zone axes weighted by their orientation fraction as defined above. The colors in the USTEM orientation map correspond to those used in the EBSD map. The USTEM map shows a gradual bend matching the sample bending found around the hole in the EBSD map of the same region. This example demonstrates potential for spatio-temporal mapping with nanoscale spatial resolution.

\begin{figure}[ht]
\centering
\includegraphics[width=0.98\linewidth]{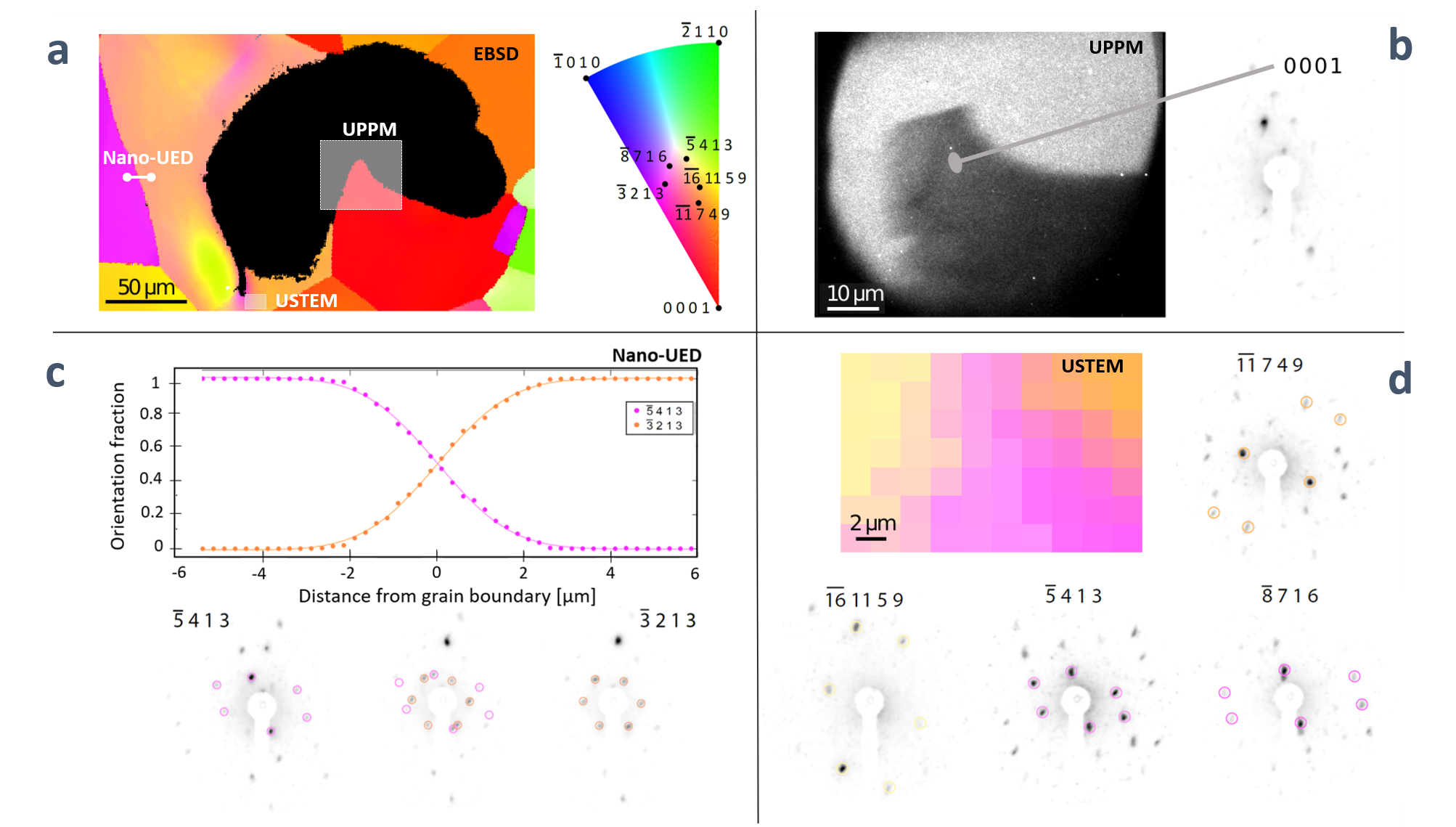}
\caption{Demonstration of correlated relativistic USTEM. \textbf{a} High resolution orientation map of a Ti-6Al polycrystal sample performed using EBSD in a scanning electron microscope. Shape and crystallographic orientation around the hole (black region in map) are used to correlate the probe position in the UED apparatus with the map. \textbf{b} UPPM of the sample in the UED setup. This location was found by searching for the highlighted feature in (a) in imaging mode and confirming the crystallographic orientation in diffraction mode. Once oriented, a scan along a chosen grain boundary (\textbf{c}) and a USTEM map (\textbf{d}) were obtained with the relativistic ultrafast electron probe.}
\label{fig:USTEM}
\end{figure}

\section*{Discussion}

We have presented experimental results demonstrating ultrafast relativistic electron scattering at the nanoscale. Electron beams as small as 90~nm have been measured, nearly two orders of magnitude smaller than previous attempts, with four-dimensional emittance values around $0.01~(nm\cdot mrad)^2$ resulting in unprecedented degree of lateral coherence. A relevant figure of merit in such regard is the relative coherence length \cite{UEDperspective_2012}, defined as $L_r=\frac{\lambda_e}{\sigma_{*}\sigma_{\theta}}$, where $\lambda_e$ is the de Broglie wavelength, $\sigma_{*}$ is the beam transverse size at the focus  and $\sigma_{\theta}$ its angular spread. From the width of the diffraction spots in Fig.~\ref{fig:USTEM} ($250~\mu m$ RMS) we obtain $\sigma_{\theta}=0.42~mrad$. Combining this information with the beam size at focus reported in Fig.\ref{roundbeam} provides 
a relative coherence length of 0.72\% and 0.43\%  for the two planes respectively, two orders of magnitude better than previously published results\cite{UCLAnanoemittance,SLAC_UED}.

Ultrafast electron-based instrumentation is not expected to have the same spatial resolution of static electron microscopes any time soon. On the other hand, ultrafast electron probes with sizes in the 100~nm range as demonstrated in this work are smaller than the typical grain size in most materials, and can therefore be used to probe local dynamics as a function of orientation or proximity to grain boundaries. As the USTEM measurement demonstrates, this technique can be applied to both sharp boundaries and graded regions, supporting spatiotemporal mapping of complex microstructures with heterogeneous orientation, composition, and phase. 
Relativistic nano-UED provides a new mean for accessing local structural information in real time, with femtosecond-nanometer resolution. The enormous scientific potential of the technique includes the study of real-time energy transfer in materials through all possible degrees of freedom, nanoscale thermal transport \cite{cahill_2003,cahill_2014}, ultrafast dynamics of individual nanowires, nanoparticles  or in low-dimensional functional nanomaterials, such as the coherent interlayer phonon excitations in epitaxial transition metal dichalcogenide (TMDC) micro-crystallites, the ultrafast manipulation of mirror domain walls in charge density wave (CDW) materials\cite{Zong2018}, and the Moir\'e pattern dynamics in unconventional superconducting magic-angle graphene superlattices\cite{Cao2018}.



We also note how there are a plethora of techniques and probes that rely on information from the microscopic and sub-microscopic world including atomic force microscopy, neutron, X-ray and electron diffraction, transmission electron microscopy and all of their variants. Often the information from one of these probes needs to be integrated and complemented with the information from the other or even better a study with one probe can be used to inform another one. This correlative microscopy can only take place if image references can be used to cross-correlate the observed position on the sample. For this reason, for example, it is essential that an instrument capable of nanodiffraction also is equipped with an imaging modality that can be effectively used in this correlation task such as the point-projection microscopy demonstrated here. 
One possible direction taking advantage of this correlative electron microscopy would be to couple high-spatial resolution maps performed at TEMs with the high temporal resolution of nano-UED setup. In this situation, the region of interest for time-resolved studies would first be selected from the TEM images, based on a specific spatial feature, and then moved to UED setups where the local dynamics can be studied. In this work, we have demonstrated how this can be enabled by UED instrumentation with both diffraction and projection microscopy capabilities, each with sub-micrometer resolution.

Lastly, these results find application also beyond the field of novel ultrafast electron scattering instrumentation in the development of ultrahigh brightness sources for injection in laser-driven micrometer aperture dielectric structures \cite{DLA}.

\section*{Methods}

\subsubsection*{UPPM contrast transfer function determination}
\label{sssec:PPIres}

We define the contrast transfer function (CTF) to be the \% of initial contrast observed for an infinite sinusoidal feature as a function of its spatial frequency $f$. Also called the modulation transfer function, this CTF can be used to compute the expected image of any feature from its spatial frequency composition \cite{boreman2001modulation}. The three-bar rulings fabricated here are finite and thus composed of many spatial frequencies; it is necessary to extract the modulation at particular spatial frequencies to determine the CTF. We base our approach on an existing procedure for three-bar gratings \cite{boreman1995modulation}. 

To prepare a ruling image for analysis, we first select a 5 pixel wide region spanning the ruling and average along its width to obtain a line profile. We then obtain a similar profile from a nearby background region, smooth with a 50 pixel Gaussian filter, and subtract the background from the ruling profile. Such a large kernel is used to avoid modifying the spatial frequency composition of the line profile in the grating frequencies of interest when subtracting the background. 

We then take the Fourier transform of the line profile to obtain its spatial frequency composition. We also take the Fourier transform of computer-generated model grating profiles having the lateral dimensions determined for the actual gratings by SEM. The modulation at a particular frequency is the ratio of the Fourier component at that frequency in the measured grating relative to that in the model grating, where both components are normalized by the component at $f=0$. 

To compare the image to the model, the image magnification must be calibrated. We take advantage of a property of three-bar gratings: the Fourier component at $\frac{f_0}{3}$ is zero. The spatial frequency at which this feature appears is not modified by applying the CTF, so we can directly determine the scale of the image from the location of the first zero point in the Fourier transform of the measured profile. We find the magnification is 145x in both directions.

For each three-bar ruling, the modulation can be most reliably determined at the fundamental frequency, $f_0$, since that is the strongest frequency component. This $f_0$ is given by the reciprocal of the grating period, which is the sum of the bar width and spacing. It is also possible to determine the modulation at other frequencies; Three-bar gratings have an  $\frac{f_0}{2}$ component that is about 1/3 the strength of the $f_0$ component. We have computed the modulation at both $f_0$ and $\frac{f_0}{2}$ for the three largest rulings as shown in Fig.~\ref{fig:PPI}c, outlining a CTF. 

We estimated the standard error in modulation measurements by simulated measurements. We consider an error dominated by random noise. We first modeled noiseless grating profiles using step functions with the lateral dimensions of the actual gratings modified by a Gaussian CTF. We then computed the noise level from a flat region of the image and added random noise at this level to the model profiles. 5000 simulated measurements were generated for each ruling by sampling the model profile with the experimental detector resolution and then adding random noise. We computed the modulation at $f_0$ and $\frac{f_0}{2}$ from all the measurements and determined the standard error represented by the error bars in Fig.~\ref{fig:PPI}c. The standard error is lowest in measurements using $f_0$ because it is the strongest frequency component. 

\subsubsection*{Relationship between UPPM resolution and beam size}
\label{sssec:PPMResToSize}

\begin{figure}[ht]
\centering
\includegraphics[width=0.6\linewidth]{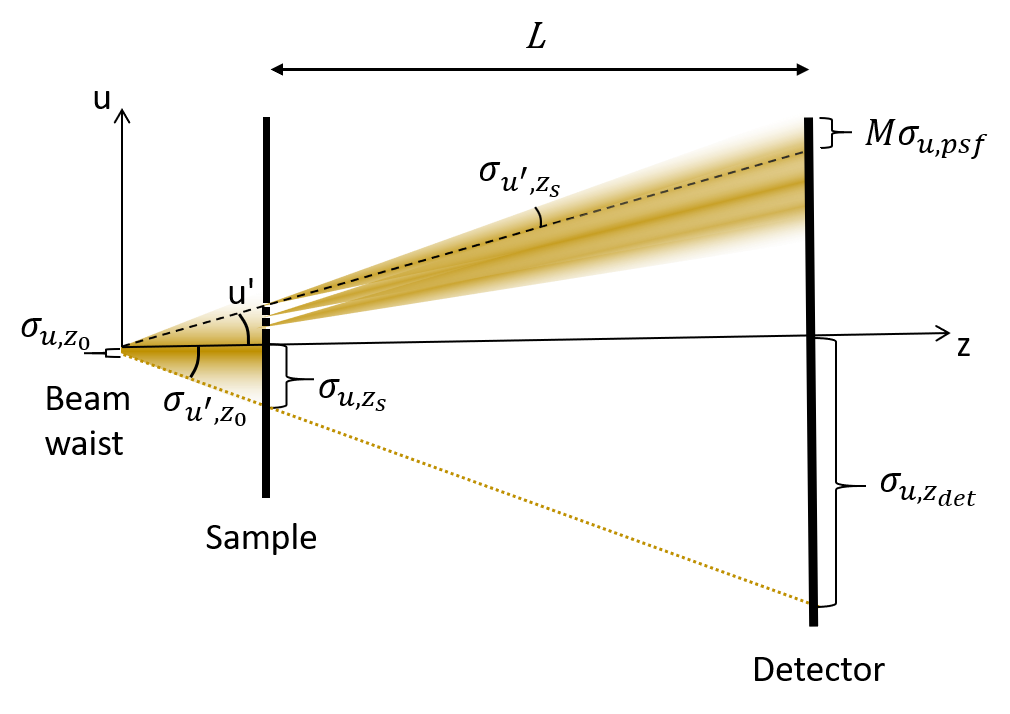}
\caption{Schematic of point-projection microscopy. This illustrates the quantities and relationships used to derive the direct relation between point spread function, $\sigma_{u,psf}$, and beam size at the waist, $\sigma_{u,z_0}$.}
\label{fig:PPMMethod}
\end{figure}

The point spread function (PSF) in point-projection microscopy is influenced by angular spread of the imaging beam (Fig.~\ref{fig:PPMMethod}). For a Gaussian beam phase space, the PSF in sample coordinates is Gaussian with $\sigma_{u,psf}=\frac{L\sigma_{u',z_s}}{M_u}$, where $\sigma_{u',z}$ is the local angle distribution, $z_s$ is the sample location, u refers to x or y coordinate, L is the distance between sample and detector, and $M_u$ is the magnification in u.

Assuming for simplicity no correlation between transverse planes, we can derive a direct relation between the PSF and the minimum beam size under the conditions depicted in Fig.~\ref{fig:PPMMethod}. We first note that the magnification $M_u=\frac{\sigma_{u,z_{det}}}{\sigma_{u,z_s}}$, with $\sigma_{u,z_{det}}$ the beam size at the detector plane and $\sigma_{u,z_s}$ the beam size at the sample plane. Also, the beam geometric emittance is constant along line (assuming no acceleration and negligible space charge forces), i.e. $\epsilon_g=\sigma_{u,z_s}\sigma_{u',z_s}=\sigma_{u,z_0}\sigma_{u',z_0}$, with $z_0$ being the beam waist position. Using these definitions and recognizing that $\sigma_{u,z_{det}} = L\sigma_{u',z_0}$, we find:

\[\sigma_{u,psf}=\frac{L\sigma_{u,z_s}\sigma_{u',z_s}}{\sigma_{u,z_{det}}}=\frac{L\sigma_{u',z_0}\sigma_{u,z_0}}{\sigma_{u,z_{det}}}=\sigma_{u,z_0}\] 

\subsubsection*{Diffraction pattern analysis}

The high resolution orientation map (Fig.~\ref{fig:USTEM}a) was acquired using scanning electron backscatter diffraction (EBSD) on an FEI Strata 235 dual beam FIB/SEM. Orientations at each position were automatically determined from the acquired EBSD patterns by the vendor's built-in pattern fitting routine. Any pattern fits with a confidence index below a threshold of 0.7 were set to black in the map: these positions are seen to correspond either to the central hole or occasionally to grain boundaries. A 3 x 3 pixel median filter was applied to the final RGB colormap to remove erroneously labeled single pixels within the large, homogeneous grains.

To identify the zone axis for each ultrafast electron diffraction pattern (DP) shown in Fig.~\ref{fig:USTEM}b-c-d, we first computed a table of reciprocal lattice basis vector pairs and corresponding zone axis based on hexagonal close-packed (HCP) Ti. We then computed the basis vector lengths and angle from each measured DP and determined the best match in the table. All patterns were confirmed by generating the expected locations of diffraction spots accounting for HCP selection rules and ensuring the measured DP satisfied these rules.

For the nano-UED line scan (Fig.~\ref{fig:USTEM}c), we used the single diffraction pattern shown from each pure grain to identify the zone axes. We only used the circled peaks to compute orientation fraction. The total diffraction signal from the grain was determined by fitting circular 2D Gaussian functions to the peaks and summing their volumes. For each zone axis, peak fitting was performed sequentially along the scan line starting from the pure grain. For the first pattern, the initial guess peak width, position, and background were set manually. For each subsequent pattern, the best-fit parameters from the previous pattern were used as the next initial guess. 

For the USTEM orientation map (Fig.~\ref{fig:USTEM}d), the grain orientation is continually rotating. We first identified the four displayed single diffraction patterns corresponding to discrete zone axes along the bend. We then estimated intermediate orientations by computing the orientation fraction of the identified discrete zone axes at each position and considering the orientation to be a linear combination of the discrete constituent zone axes. This allows us to reproduce the bending feature from the EBSD map (Fig.~\ref{fig:USTEM}a). In this case, the initial guess for peak parameters for a given orientation at each position used the best-fit values from the corresponding representative single diffraction pattern. 

\bibliography{main}

\begin{thebibliography}{10}
\expandafter\ifx\csname url\endcsname\relax
  \def\url#1{\texttt{#1}}\fi
\expandafter\ifx\csname urlprefix\endcsname\relax\def\urlprefix{URL }\fi
\expandafter\ifx\csname doiprefix\endcsname\relax\def\doiprefix{DOI }\fi
\providecommand{\bibinfo}[2]{#2}
\providecommand{\eprint}[2][]{\url{#2}}

\bibitem{Nobel}
\bibinfo{title}{The nobel prize in physics 1937}.
\newblock
  \bibinfo{howpublished}{\url{http://www.nobelprize.org/nobel_prizes/physics/laureates/1937/}}.

\bibitem{Zewail}
\bibinfo{author}{Zewail, A.~H.}
\newblock \bibinfo{journal}{\bibinfo{title}{4d ultrafast electron diffraction,
  crystallography, and microscopy}}.
\newblock {\emph{\JournalTitle{Annu. Rev. Phys. Chem.}}}
  \textbf{\bibinfo{volume}{57}}, \bibinfo{pages}{65--103}
  (\bibinfo{year}{2006}).

\bibitem{SciainiMiller}
\bibinfo{author}{Sciaini, G.} \& \bibinfo{author}{Miller, R.~D.}
\newblock \bibinfo{journal}{\bibinfo{title}{Femtosecond electron diffraction:
  heralding the era of atomically resolved dynamics}}.
\newblock {\emph{\JournalTitle{Reports on Progress in Physics}}}
  \textbf{\bibinfo{volume}{74}}, \bibinfo{pages}{096101}
  (\bibinfo{year}{2011}).

\bibitem{king_JAP2005}
\bibinfo{author}{King, W.~E.} \emph{et~al.}
\newblock \bibinfo{journal}{\bibinfo{title}{Ultrafast electron microscopy in
  materials science, biology, and chemistry}}.
\newblock {\emph{\JournalTitle{Journal of Applied Physics}}}
  \textbf{\bibinfo{volume}{97}}, \bibinfo{pages}{8} (\bibinfo{year}{2005}).

\bibitem{maxson2017}
\bibinfo{author}{Maxson, J.} \emph{et~al.}
\newblock \bibinfo{journal}{\bibinfo{title}{Direct measurement of sub-10 fs
  relativistic electron beams with ultralow emittance}}.
\newblock {\emph{\JournalTitle{Physical review letters}}}
  \textbf{\bibinfo{volume}{118}}, \bibinfo{pages}{154802}
  (\bibinfo{year}{2017}).

\bibitem{zhao2018}
\bibinfo{author}{Zhao, L.} \emph{et~al.}
\newblock \bibinfo{journal}{\bibinfo{title}{Few-femtosecond electron beam with
  terahertz-frequency wakefield-driven compression}}.
\newblock {\emph{\JournalTitle{Physical Review Accelerators and Beams}}}
  \textbf{\bibinfo{volume}{21}}, \bibinfo{pages}{082801}
  (\bibinfo{year}{2018}).

\bibitem{SLAC_first}
\bibinfo{author}{Hastings, J.~B.} \emph{et~al.}
\newblock \bibinfo{journal}{\bibinfo{title}{Ultrafast time-resolved electron
  diffraction with megavolt electron beams}}.
\newblock {\emph{\JournalTitle{Applied Physics Letters}}}
  \textbf{\bibinfo{volume}{89}}, \bibinfo{pages}{184109}
  (\bibinfo{year}{2006}).
\newblock \doiprefix 10.1063/1.2372697.

\bibitem{yang2016}
\bibinfo{author}{Yang, J.} \emph{et~al.}
\newblock \bibinfo{journal}{\bibinfo{title}{Diffractive imaging of a rotational
  wavepacket in nitrogen molecules with femtosecond megaelectronvolt electron
  pulses}}.
\newblock {\emph{\JournalTitle{Nature communications}}}
  \textbf{\bibinfo{volume}{7}}, \bibinfo{pages}{ncomms11232}
  (\bibinfo{year}{2016}).

\bibitem{BazarovPRL}
\bibinfo{author}{Bazarov, I.~V.}, \bibinfo{author}{Dunham, B.~M.} \&
  \bibinfo{author}{Sinclair, C.~K.}
\newblock \bibinfo{journal}{\bibinfo{title}{Maximum achievable beam brightness
  from photoinjectors}}.
\newblock {\emph{\JournalTitle{Phys. Rev. Lett.}}}
  \textbf{\bibinfo{volume}{102}}, \bibinfo{pages}{104801}
  (\bibinfo{year}{2009}).
\newblock \doiprefix 10.1103/PhysRevLett.102.104801.

\bibitem{filippetto2014}
\bibinfo{author}{Filippetto, D.}, \bibinfo{author}{Musumeci, P.},
  \bibinfo{author}{Zolotorev, M.} \& \bibinfo{author}{Stupakov, G.}
\newblock \bibinfo{journal}{\bibinfo{title}{Maximum current density and beam
  brightness achievable by laser-driven electron sources}}.
\newblock {\emph{\JournalTitle{Physical Review Special Topics-Accelerators and
  Beams}}} \textbf{\bibinfo{volume}{17}}, \bibinfo{pages}{024201}
  (\bibinfo{year}{2014}).

\bibitem{ropers2014}
\bibinfo{author}{Gulde, M.} \emph{et~al.}
\newblock \bibinfo{journal}{\bibinfo{title}{Ultrafast low-energy electron
  diffraction in transmission resolves polymer/graphene superstructure
  dynamics}}.
\newblock {\emph{\JournalTitle{Science}}} \textbf{\bibinfo{volume}{345}},
  \bibinfo{pages}{200--204} (\bibinfo{year}{2014}).

\bibitem{Arbouet2017}
\bibinfo{author}{Caruso, G.~M.}, \bibinfo{author}{Houdellier, F.},
  \bibinfo{author}{Abeilhou, P.} \& \bibinfo{author}{Arbouet, A.}
\newblock \bibinfo{journal}{\bibinfo{title}{Development of an ultrafast
  electron source based on a cold-field emission gun for ultrafast coherent
  {TEM}}}.
\newblock {\emph{\JournalTitle{Applied Physics Letters}}}
  \textbf{\bibinfo{volume}{111}}, \bibinfo{pages}{023101}
  (\bibinfo{year}{2017}).
\newblock \doiprefix 10.1063/1.4991681.

\bibitem{SLAC_UED}
\bibinfo{author}{Weathersby, S.} \emph{et~al.}
\newblock \bibinfo{journal}{\bibinfo{title}{Mega-electron-volt ultrafast
  electron diffraction at slac national accelerator laboratory}}.
\newblock {\emph{\JournalTitle{Review of Scientific Instruments}}}
  \textbf{\bibinfo{volume}{86}}, \bibinfo{pages}{073702}
  (\bibinfo{year}{2015}).

\bibitem{BNL_UED}
\bibinfo{author}{Zhu, P.} \emph{et~al.}
\newblock \bibinfo{journal}{\bibinfo{title}{Femtosecond time-resolved {MeV}
  electron diffraction}}.
\newblock {\emph{\JournalTitle{New Journal of Physics}}}
  \textbf{\bibinfo{volume}{17}}, \bibinfo{pages}{063004}
  (\bibinfo{year}{2015}).
\newblock \doiprefix 10.1088/1367-2630/17/6/063004.

\bibitem{sannibale2012}
\bibinfo{author}{Sannibale, F.} \emph{et~al.}
\newblock \bibinfo{journal}{\bibinfo{title}{Advanced photoinjector experiment
  photogun commissioning results}}.
\newblock {\emph{\JournalTitle{Physical Review Special Topics-Accelerators and
  Beams}}} \textbf{\bibinfo{volume}{15}}, \bibinfo{pages}{103501}
  (\bibinfo{year}{2012}).

\bibitem{mullerPPM2014}
\bibinfo{author}{MÃŒller, M.}, \bibinfo{author}{Paarmann, A.} \&
  \bibinfo{author}{Ernstorfer, R.}
\newblock \bibinfo{journal}{\bibinfo{title}{Femtosecond electrons probing
  currents and atomic structure in nanomaterials}}.
\newblock {\emph{\JournalTitle{Nature Communications}}}
  \textbf{\bibinfo{volume}{5}} (\bibinfo{year}{2014}).
\newblock \doiprefix 10.1038/ncomms6292.

\bibitem{bainbridgeUPPM2016}
\bibinfo{author}{Bainbridge, A.~R.}, \bibinfo{author}{Barlow~Myers, C.~W.} \&
  \bibinfo{author}{Bryan, W.~A.}
\newblock \bibinfo{journal}{\bibinfo{title}{Femtosecond few- to single-electron
  point-projection microscopy for nanoscale dynamic imaging}}.
\newblock {\emph{\JournalTitle{Structural Dynamics}}}
  \textbf{\bibinfo{volume}{3}}, \bibinfo{pages}{023612} (\bibinfo{year}{2016}).
\newblock \doiprefix 10.1063/1.4947098.

\bibitem{quinonez_UPPM2013}
\bibinfo{author}{Quinonez, E.}, \bibinfo{author}{Handali, J.} \&
  \bibinfo{author}{Barwick, B.}
\newblock \bibinfo{journal}{\bibinfo{title}{Femtosecond photoelectron point
  projection microscope}}.
\newblock {\emph{\JournalTitle{Review of Scientific Instruments}}}
  \textbf{\bibinfo{volume}{84}}, \bibinfo{pages}{103710}
  (\bibinfo{year}{2013}).
\newblock \doiprefix 10.1063/1.4827035.

\bibitem{DCdesign_1953}
\bibinfo{author}{Cowley, J.~M.} \& \bibinfo{author}{Rees, A. L.~G.}
\newblock \bibinfo{journal}{\bibinfo{title}{Design of a high-resolution
  electron diffraction camera}}.
\newblock {\emph{\JournalTitle{Journal of Scientific Instruments}}}
  \textbf{\bibinfo{volume}{30}}, \bibinfo{pages}{33--38}
  (\bibinfo{year}{1953}).
\newblock \doiprefix 10.1088/0950-7671/30/2/301.

\bibitem{UCLAnanoemittance}
\bibinfo{author}{Li, R.~K.}, \bibinfo{author}{Roberts, K.~G.},
  \bibinfo{author}{Scoby, C.~M.}, \bibinfo{author}{To, H.} \&
  \bibinfo{author}{Musumeci, P.}
\newblock \bibinfo{journal}{\bibinfo{title}{Nanometer emittance ultralow charge
  beams from rf photoinjectors}}.
\newblock {\emph{\JournalTitle{Phys. Rev. ST Accel. Beams}}}
  \textbf{\bibinfo{volume}{15}}, \bibinfo{pages}{090702}
  (\bibinfo{year}{2012}).
\newblock \doiprefix 10.1103/PhysRevSTAB.15.090702.

\bibitem{HiRESsims}
\bibinfo{author}{Filippetto, D.} \& \bibinfo{author}{Qian, H.}
\newblock \bibinfo{journal}{\bibinfo{title}{Design of a high-flux instrument
  for ultrafast electron diffraction and microscopy}}.
\newblock {\emph{\JournalTitle{Journal of Physics B: Atomic, Molecular and
  Optical Physics}}} \textbf{\bibinfo{volume}{49}}, \bibinfo{pages}{104003}
  (\bibinfo{year}{2016}).

\bibitem{APEXgun}
\bibinfo{author}{Sannibale, F.} \emph{et~al.}
\newblock \bibinfo{journal}{\bibinfo{title}{Advanced photoinjector experiment
  photogun commissioning results}}.
\newblock {\emph{\JournalTitle{Physical Review Special Topics - Accelerators
  and Beams}}} \textbf{\bibinfo{volume}{15}} (\bibinfo{year}{2012}).
\newblock \doiprefix 10.1103/PhysRevSTAB.15.103501.

\bibitem{MarxNavarro_prab}
\bibinfo{author}{Marx, D.} \emph{et~al.}
\newblock \bibinfo{journal}{\bibinfo{title}{Single-shot reconstruction of core
  4d phase space of high-brightness electron beams using metal grids}}.
\newblock {\emph{\JournalTitle{Physical Review Accelerators and Beams}}}
  \textbf{\bibinfo{volume}{21}}, \bibinfo{pages}{102802}
  (\bibinfo{year}{2018}).

\bibitem{Nanoquadrupoles1}
\bibinfo{author}{Lim, J.~K.} \emph{et~al.}
\newblock \bibinfo{journal}{\bibinfo{title}{Adjustable, short focal length
  permanent-magnet quadrupole based electron beam final focus system}}.
\newblock {\emph{\JournalTitle{Physical Review Special Topics - Accelerators
  and Beams}}} \textbf{\bibinfo{volume}{8}} (\bibinfo{year}{2005}).
\newblock \doiprefix 10.1103/PhysRevSTAB.8.072401.

\bibitem{Nanoquadrupoles2}
\bibinfo{author}{Li, R.} \& \bibinfo{author}{Musumeci, P.}
\newblock \bibinfo{journal}{\bibinfo{title}{Single-{Shot} {MeV} {Transmission}
  {Electron} {Microscopy} with {Picosecond} {Temporal} {Resolution}}}.
\newblock {\emph{\JournalTitle{Physical Review Applied}}}
  \textbf{\bibinfo{volume}{2}} (\bibinfo{year}{2014}).
\newblock \doiprefix 10.1103/PhysRevApplied.2.024003.

\bibitem{GPT}
\bibinfo{title}{General particle tracer}.
\newblock \bibinfo{howpublished}{\url{http://www.pulsar.nl/gpt}}.

\bibitem{PSI_nanobeam}
\bibinfo{author}{Borrelli, S.} \emph{et~al.}
\newblock \bibinfo{journal}{\bibinfo{title}{Generation and measurement of
  sub-micrometer relativistic electron beams}}.
\newblock {\emph{\JournalTitle{Communications Physics}}}
  \textbf{\bibinfo{volume}{1}} (\bibinfo{year}{2018}).
\newblock \doiprefix 10.1038/s42005-018-0048-x.

\bibitem{KnifeEdge_emitt}
\bibinfo{author}{Ji, F.} \emph{et~al.}
\newblock \bibinfo{journal}{\bibinfo{title}{in preparation}}.
\newblock {\emph{\JournalTitle{2018}}} .

\bibitem{ronchi1923frange}
\bibinfo{author}{Ronchi, V.}
\newblock \bibinfo{journal}{\bibinfo{title}{Le frange di combinazioni nello
  studio delle superficie e dei sistemi ottici}}.
\newblock {\emph{\JournalTitle{Riv. Ottica Mecc. Precis}}}
  \textbf{\bibinfo{volume}{2}} (\bibinfo{year}{1923}).

\bibitem{osterberg1962evaluation}
\bibinfo{author}{Osterberg, H.}
\newblock \bibinfo{journal}{\bibinfo{title}{Evaluation phase optical tests}}.
\newblock {\emph{\JournalTitle{Military Standardization Handbook: Optical
  Design}}} \bibinfo{pages}{1--8} (\bibinfo{year}{1962}).

\bibitem{UEDperspective_2012}
\bibinfo{author}{Carbone, F.}, \bibinfo{author}{Musumeci, P.},
  \bibinfo{author}{Luiten, O.} \& \bibinfo{author}{Hebert, C.}
\newblock \bibinfo{journal}{\bibinfo{title}{A perspective on novel sources of
  ultrashort electron and {X}-ray pulses}}.
\newblock {\emph{\JournalTitle{Chemical Physics}}}
  \textbf{\bibinfo{volume}{392}}, \bibinfo{pages}{1--9} (\bibinfo{year}{2012}).
\newblock \doiprefix 10.1016/j.chemphys.2011.10.010.

\bibitem{cahill_2003}
\bibinfo{author}{Cahill, D.~G.} \emph{et~al.}
\newblock \bibinfo{journal}{\bibinfo{title}{Nanoscale thermal transport}}.
\newblock {\emph{\JournalTitle{Journal of Applied Physics}}}
  \textbf{\bibinfo{volume}{93}}, \bibinfo{pages}{793--818}
  (\bibinfo{year}{2003}).
\newblock \doiprefix 10.1063/1.1524305.

\bibitem{cahill_2014}
\bibinfo{author}{Cahill, D.~G.} \emph{et~al.}
\newblock \bibinfo{journal}{\bibinfo{title}{Nanoscale thermal transport. {II}.
  2003-2012}}.
\newblock {\emph{\JournalTitle{Applied Physics Reviews}}}
  \textbf{\bibinfo{volume}{1}}, \bibinfo{pages}{011305} (\bibinfo{year}{2014}).
\newblock \doiprefix 10.1063/1.4832615.

\bibitem{Zong2018}
\bibinfo{author}{Zong, A.} \emph{et~al.}
\newblock \bibinfo{journal}{\bibinfo{title}{Ultrafast manipulation of mirror
  domain walls in a charge density wave}}.
\newblock {\emph{\JournalTitle{Science Advances}}}
  \textbf{\bibinfo{volume}{4}}, \bibinfo{pages}{5501} (\bibinfo{year}{2018}).

\bibitem{Cao2018}
\bibinfo{author}{Cao, Y.} \emph{et~al.}
\newblock \bibinfo{journal}{\bibinfo{title}{Unconventional superconductivity in
  magic-angle graphene superlattices}}.
\newblock {\emph{\JournalTitle{Nature}}} \textbf{\bibinfo{volume}{556}},
  \bibinfo{pages}{43–50} (\bibinfo{year}{2018}).

\bibitem{DLA}
\bibinfo{author}{England, R.~J.} \emph{et~al.}
\newblock \bibinfo{journal}{\bibinfo{title}{Dielectric laser accelerators}}.
\newblock {\emph{\JournalTitle{Reviews of Modern Physics}}}
  \textbf{\bibinfo{volume}{86}}, \bibinfo{pages}{1337--1389}
  (\bibinfo{year}{2014}).
\newblock \doiprefix 10.1103/RevModPhys.86.1337.

\bibitem{boreman2001modulation}
\bibinfo{author}{Boreman, G.~D.}
\newblock \emph{\bibinfo{title}{Modulation transfer function in optical and
  electro-optical systems}}, vol.~\bibinfo{volume}{21}
  (\bibinfo{publisher}{SPIE press Bellingham, WA}, \bibinfo{year}{2001}).

\bibitem{boreman1995modulation}
\bibinfo{author}{Boreman, G.~D.} \& \bibinfo{author}{Yang, S.}
\newblock \bibinfo{journal}{\bibinfo{title}{Modulation transfer function
  measurement using three-and four-bar targets}}.
\newblock {\emph{\JournalTitle{Applied optics}}} \textbf{\bibinfo{volume}{34}},
  \bibinfo{pages}{8050--8052} (\bibinfo{year}{1995}).

\end{thebibliography}

\section*{Acknowledgements}

The authors would like to thank D. Arbelaez, T. Luo and the Berkeley Center For Magnet Technology, for providing help and tools for magnetic measurements. DF and FJ were supported by the U.S. Department of Energy under Contract No. DE-AC02-05CH11231. The experimental chamber hardware necessary for the experiment was acquired thanks to funds provided by LBNL through the Laboratory Directed Research and Development plan (LDRD). This LBNL-UCB-UCLA collaboration was supported by STROBE: A National Science Foundation Science and Technology Center under Grant No. DMR 1548924, which provided funding for DD. JGN acknowledges support from the Center for Bright Beams, NSF award PHY 1549132. We thank R. Zhang at UC Berkeley for help preparing the Ti-Al sample and EBSD map. EBSD and all sample preparation was supported by the Molecular Foundry at Lawrence Berkeley National Laboratory, which is supported by the U.S. Department of Energy under Contract DE-AC02-05CH11231. 

\section*{Author contributions statement}

DF conceived and designed the experiment. DF, PM and FJ performed electron beam simulations of the beamline. DF, FJ and DD conducted the experiments. DD and AM prepared and characterized the samples for knife-edge and USTEM measurements. FJ, JN, PM, DD and DF analyzed the results. All authors reviewed the manuscript.

\section*{Additional information}

\textbf{Competing financial interests:} The authors declare no competing interests.


\end{document}